\documentclass[a4paper]{article}

\usepackage{INTERSPEECH2020}
\usepackage{amsmath,amssymb,graphicx, dsfont, subcaption, adjustbox}
\usepackage{color}
\usepackage{cite}
\usepackage{balance}

\usepackage[ruled, linesnumbered]{algorithm2e}
\usepackage{wasysym}
\usepackage{bm}
\usepackage{booktabs}
\usepackage{enumitem} %
\usepackage{pifont}%

\def\RR{\mathbb R}

\DeclareMathOperator*{\argmin}{argmin}

\def\vec#1{\ensuremath{\bm{{#1}}}}

\title{Detecting Audio Attacks on ASR Systems with Dropout Uncertainty}
\name{Tejas Jayashankar$^{1,2}$, Jonathan Le Roux$^1$, Pierre Moulin$^{1,2}$\thanks{This work was performed while T.~Jayashankar was an intern and P.~Moulin on sabbatical at MERL.}}
\address{$^1$Mitsubishi Electric Research Laboratories (MERL), Cambridge, MA, USA\\
$^2$University of Illinois at Urbana-Champaign, Champaign, IL, USA}
\email{tejaskj@gmail.com,leroux@merl.com,moulin@ifp.uiuc.edu}

\begin{document}
\maketitle
\setlength{\abovedisplayskip}{4pt}
\setlength{\belowdisplayskip}{4pt}

\begin{abstract}
Various adversarial audio attacks have recently been developed to fool automatic speech recognition (ASR) systems. We here propose a defense against such attacks based on the uncertainty introduced by dropout in neural networks. We show that our defense is able to detect attacks created through optimized perturbations and frequency masking on a state-of-the-art end-to-end ASR system. Furthermore, the defense can be made robust against attacks that are immune to noise reduction. We test our defense on Mozilla's CommonVoice dataset, the UrbanSound dataset, and an excerpt of the LibriSpeech dataset, showing that it achieves high detection accuracy in a wide range of scenarios.
\end{abstract}

\section{Introduction}
An adversarial example is an input to a neural network designed by an adversary to produce an incorrect or malicious output \cite{Biggio2013}.  Early work on adversarial machine learning has shown that a small and imperceptible optimized perturbation to an image can cause misclassification by neural networks \cite{Szegedy2013}.  The field has further expanded to tasks such as image segmentation \cite{Arnab2017}, reinforcement learning \cite{Huang2017}, and reading comprehension \cite{Jia2017}.  

Recently, there has been growing interest in creating adversarial audio examples for automatic speech recognition (ASR) systems.  Carlini and Wagner \cite{Carlini2018} showed that an audio sample can be perturbed slightly to cause mistranscription by an ASR engine.  Building on this model, Qin et al.\ \cite{Qin2019} and Sch{\"o}nherr et al.\ \cite{schonherr2018adversarial,schonherr2019robust} created nearly imperceptible audio attacks by leveraging the principle of auditory masking \cite{masking-book}. There have also been attempts to create attacks embedded in ultrasound frequencies \cite{Zhang2017} as well as phonetically constrained attacks \cite{Ciss2017}.

Adversarial machine learning exploits a vulnerability of neural network models but also provides an avenue for making models more robust and formulating defenses against such attacks.  There has been a significant effort in understanding the underlying mechanism of adversarial attacks to formulate effective defenses against attacks \cite{Carlini2017, Tramr2017}, however such work has largely focused on domains other than audio. %

In this paper, we propose a defense against adversarial audio attacks based on dropout. Dropout \cite{Srivastava2014} is heavily used as an effective regularizer for neural network training, particularly in ASR systems. There have been successful attempts at using dropout as a defense in the image domain \cite{Feinman2017}. We here investigate whether similar principles can be applied to ASR, where varying sequence lengths pose an additional challenge. While the analysis of discrepancies in dropout outputs has been investigated to model uncertainty in ASR hypotheses \cite{vyas2019analyzing}, to our knowledge this is its first use as a defense against audio attacks.

\section{End-to-end automatic speech recognition}
Many recent ASR systems obtaining state-of-the-art results are based on end-to-end architectures \cite{Prabhavalkar2017}.
In contrast with conventional hybrid ASR systems, which consist in multiple complex modules such as acoustic, lexicon, and language models, end-to-end systems typically use a single deep network trained to directly map an input audio sample to a sequence of words or characters, alleviating the need for expert knowledge to build competitive systems.

The most popular end-to-end ASR approaches are connectionist temporal classification (CTC) \cite{Graves2006, Hannun2014}, attention \cite{Chan2016}, CTC/attention \cite{Kim2016}, RNN-T \cite{Graves2013}, and the Transformer \cite{Zhou2018,karita2019comparative}.  Since these models are differentiable, they can be trained with backpropagation, which is appealing due to the ease with which the model parameters can be updated by employing the chain rule of differentiation.  However, this can also be a weakness, because an adversary may craft an adversarial input to fool a model into producing a wrong or a malicious output by backpropagating through it in order to minimize an error loss between the output of the model and the desired output. %

In this paper, we focus on a CTC based architecture, as implemented in Mozilla's DeepSpeech system \cite{Hannun2014}. That system has been used in past work on adversarial audio attacks \cite{Carlini2018} and is publicly available, making it a convenient subject for our study.  However, the methods we detail can be applied to models based on other end-to-end architectures as well.

\section{Dropout}
\label{sec:dropout}

Dropout \cite{Srivastava2014} is a regularization technique used to make neural networks robust to different inputs.  Dropout deactivates a certain number of neurons in a layer, i.e., the weights corresponding to the neurons are set to zero.  In each training iteration, a layer with dropout rate $p$ drops neurons uniformly at random with probability $p$. %
During inference, dropout is typically turned off, and the learned weight matrices are scaled by $p$ so that the expected value of an activation is the same as during training. Intuitively, dropout enables the neural network to learn various internal representations for the same input and output pair. 

Adversaries typically exploit loopholes within a network by crafting an input perturbation such that small finely-tuned differences accumulate within the network to eventually result in a malicious output.  Since these adversarial attacks are often created based on knowledge of the underlying architecture of the model, we hope to disarm such attacks by perturbing that architecture via a random process like dropout.

\section{Adversarial attacks on ASR systems}
\label{sec:attacks}

In this section, we present the various adversarial audio attacks that we use in our experiments.  We first introduce the Carlini \& Wagner attack (CW attack) as it forms the foundation for the other attacks that we consider.

\subsection{Carlini \& Wagner Attack}
\label{sec:CW-attack}

Given an original waveform $x$, Carlini and Wagner \cite{Carlini2018} propose to construct a waveform $x^\prime = x + \delta$ such that $x$ and $x^\prime$ sound nearly the same but are transcribed differently by an ASR engine. %
The perturbation $\delta$ is optimized such that the perturbed waveform $x+\delta$ is transcribed as a specific alternate (typically malicious) target sentence $t$ with the least distortion, by minimizing a recognition loss function $\ell(x+\delta,t)$ (here, a CTC loss) under the constraint that the peak energy of the perturbation be at least $\tau$~dB smaller than that of the original waveform:
\begin{align}
    \min_\delta \ell(x+\delta, t)\text{ s.t. }\mathrm{dB}(\delta) \le \mathrm{dB}(x)-\tau,
    \label{eq:CW_optim}
\end{align}
where $\mathrm{dB}(x) = 20 \max_i \log(|x_i|)$. Because $\ell$ is differentiable, backpropagation is easily performed. During optimization, the values of $\delta$ are limited to avoid clipping, and the threshold $\tau$ is progressively decreased to strengthen the constraint. %

\subsection{Dropout Robust Attack}
\label{sec:dropout-robust}

Adversarial examples generated by the CW attack are optimized to be transcribed as certain target sentences at test time, with the model in inference mode, and they are thus typically optimized through the model with dropout turned off. A key insight is that if inference is performed with %
dropout turned on, adversarial examples tend to be transcribed as incorrect or garbled sentences, and dropout may thus be used to detect them. %
But we may consider a dropout robust (DR) attack including dropout in the construction of the adversarial example, so that the optimization procedure %
may have the chance to account for it.
Since the adversarial example should be transcribed as target sentence $t$ both with and without dropout turned on at inference time, the loss in (\ref{eq:CW_optim}) is replaced by a multi-task loss formulated as
\begin{align}
    \label{eq:dropout_optim}
    \min_\delta \ell(x\!+\!\delta, t) \!+\! \beta \ell_{p_\text{DR}}(x\!+\!\delta, t)
     \text{ s.t. }\mathrm{dB}(\delta) \!\le\!  \mathrm{dB}(x) - \tau,
\end{align}
where %
$\ell_{p_\text{DR}}(x+\delta,t)$ is the same loss as $\ell$ except that dropout is turned on with a rate $p_\text{DR}$, and $\beta$ is a weight (we used $\beta=1$). %
Dropout is applied to all layers except the LSTM layers. %

\subsection{Noise Reduction Robust (NRR) Attack}
\label{sec:noise-reduction-robust}

Many audio systems perform pre-processing steps involving denoising to clean the input audio signal.  We observed that denoising could partially and often completely eliminate the perturbation in the CW adversarial samples. Thus, denoising could act in itself as an effective defense for the vanilla CW attack. 

To make the attack more robust, we trained the adversary to transcribe as the target sentence $t$ with and without a pre-processing denoising stage by backpropagating through spectral subtraction \cite{Boll1979}, which was chosen for its simplicity: %
\begin{align}
    \label{eq:noisered_optim}
    \min_\delta  \ell(x\!+\!\delta, t) \!+\! \beta \ell_{\mathrm{ss}}(x\!+\!\delta, t)
     \text{ s.t. }\mathrm{dB}(\delta) \!\le\!  \mathrm{dB}(x) - \tau,
\end{align}
where %
$\ell_{\mathrm{ss}}(x+\delta,t)$ is the same loss as $\ell$ except that the network processes %
the perturbed input after spectral subtraction denoising. %
We also tried to make this attack robust to dropout as above, but the optimization failed to converge in a reasonable time, illustrating the difficulty to find a solution under such constraints. %

We did not experiment with neural network denoising algorithms as these could themselves be susceptible to adversarial attacks through similar optimization procedures. %

\subsection{Imperceptible Audio Attack}
\label{sec:impercptible-audio-attack}

Recently, Qin et al.\ \cite{Qin2019} devised a new attack on ASR systems based on frequency masking, the phenomenon whereby a softer sound (the maskee) is rendered inaudible by a louder sound (the masker) \cite{masking-book}.  
The vanilla CW attack is modified to enforce that the power spectral density $p_\delta$ of the perturbation in the short-time Fourier transform (STFT) domain must fall below the masking threshold $\theta_x$ of the original audio sample.  The complete optimization problem is formulated as
\begin{equation}
    \min_\delta \ell(x+\delta, t) + \alpha \sum_{k=0}^{\lfloor \frac{N}{2}\rfloor} \max\{p_\delta(k) - \theta_x(k), 0\},
\end{equation}
where %
$\alpha$ controls the relative importance of the term making the perturbation imperceptible, and $N$ is the STFT window size. %

After first optimizing with $\alpha=0$ to find a perturbed sample transcribing as $t$, $\alpha$ is slowly increased to gradually satisfy the imperceptibility constraint by fine-tuning the perturbation.

\subsection{Urban Sound Attack}
\label{sec:urban-sound-attack}

We  also apply the vanilla CW attack to audio recordings of every day noises such as construction sounds, cars honking, and leaves rustling.  The aim of this experiment is two-fold: 1) Can the vanilla CW attack be applied to general sounds? 2) Can our defense detect attacks concealed in such audio recordings?
\subsection{Universal Perturbation Attack}
\label{sec:universal-attack}

Finally, we study adversarial examples generated by a model based on universal adversarial perturbations \cite{Neekhara2019}. A universal perturbation is a single perturbation which when added to different input audio samples causes a mistranscription by the ASR engine.  Unlike the perturbations considered so far, universal perturbations are not targeted attacks, i.e., the transcription produced is not fixed. Moreover,  in most cases, the transcription is not a meaningful sentence. %

\section{Proposed Defense}
\label{sec:our-defense}
\subsection{Dropout defense in the image domain}
Feinman et al.\ \cite{Feinman2017} showed that dropout can be used to build an uncertainty estimator in neural networks for image classification.  Specifically, dropout in neural networks mimics a deep Gaussian process and hence Bayesian estimates can be inferred.  In their experiments, they subject an input image to $I$ realizations of dropout during inference. The intuition is that the realizations obtained from an adversarial example will show more variation than those obtained from an original example. 
Let us denote by $y(x, \mathbf{W})\!\in\! \RR^C$ the output probability vector of an image classification network with parameters $\mathbf{W}$ for an input image $x$, where $C$ denotes the number of classes.
Each realization of dropout results in a new set of network parameters $\mathbf{W}^{(i)},\, i = 1,\dots, I$. %
The output for realization $i$ %
is denoted as
\begin{equation}
    \label{eq:network-output}
    y_i = y(x, \mathbf{W}^{(i)}).
\end{equation}
The uncertainty $U(x)$ of the network with respect to input $x$ is defined as the trace of the covariance matrix of the realizations, or equivalently as the average Euclidean distance between the realizations and their mean $\hat{y} = \frac{1}{I}\sum_{i=1}^I y_i$:
\begin{equation}
     U(x) = \dfrac{1}{I}\sum_{t=1}^I \| y_i - \hat{y}\|^2.
    \label{eq:uncertainty-dropout-image}
\end{equation}
A simple threshold-based classifier using the scalar $U(x)$ as input can now be designed to classify original and adversarial samples, as the uncertainty of adversarial samples is expected to be higher than that of original samples on average.

\subsection{Extending the notion of uncertainty}
\label{sec:uncertainty}
Before we move to the audio domain, let us first introduce a generalization to the notion of uncertainty used in Feinman et al.\ \cite{Feinman2017}, which will be useful later on. Instead of a single number, we would like to extract richer features for classification. We assume we have a set $\{y_i\}_{i=1}^I$ of $I$ points in some space $X$, obtained as realizations of a neural network output with dropout. We also assume that we have some function $d$ measuring a notion of distance between two points in $X$, as well as a mechanism to obtain a point $\hat{y}$ from the set $\{y_i\}_{i=1}^I$ encompassing some notion of average with respect to these points. In the image classification case above, the space $X$ is the Euclidean space $\RR^C$, the function $d$ is the squared Euclidean distance, and $\hat{y}$ is obtained as the mean of the points in $\{y_i\}_{i=1}^I$. Based on these components, we can define the uncertainty distribution  
\begin{equation}
\label{eq:uncertainty_distribution}
\mathds{P}(z) = \sum_{i} \mathds{1}_{\{d(\hat{y}, y_i) = z\}}, \, z \in \RR^+,
\end{equation}
from which we can extract features to be used by a classifier. For instance, in the image case, the uncertainty $U(x)$ of Eq.~\eqref{eq:uncertainty-dropout-image} is none other than the second moment of $\mathds{P}$. %

\subsection{Designing a notion of uncertainty for ASR}
\label{sec:designing}
The defense devised by Feinman cannot be directly applied to the ASR case. Indeed, contrary to image classification where the network output is a vector of class probability predictions with fixed length, %
the corresponding output of an ASR system, in the case of CTC, is a sequence of such posterior probabilities, whose length depends on the input length. The problem is even more complex for decoder-based models, where the length of that sequence also depends on the internal processing of the network, as the decoder determines itself the output length. 

A direct extension of Feinman's defense to the ASR case could be to consider the sequence of CTC posterior probabilities for an input $x$ as a large vector used as the realization $y_i$, and to compute the uncertainty as in Eq.~\eqref{eq:uncertainty-dropout-image}, normalizing by the input length. We consider this our baseline defense. We can further generalize this defense by considering the uncertainty distribution $\mathds{P}_x^\text{prob}$ obtained using Eq.~\eqref{eq:uncertainty_distribution} in this context, and deriving features from it for a classifier.

For greater generalizability to various architectures and to reduce the dependence on the audio input length, we consider designing a defense based not on the sequence of CTC posterior probabilities but on the final output character sequence, which stems from all components of the network, including a potential language model. %
The final character sequence length for a given input may however vary depending on the internal processing of the network. Transcriptions for different dropout realizations may thus be of different lengths.
Furthermore, while probability vectors can be considered within a Euclidean space, %
this is not possible for character sequences. 

To define an uncertainty distribution following Section~\ref{sec:uncertainty},
we thus need to use a (non-Euclidean) distance metric $d$ that can be calculated between character sequences with potentially different lengths. 
We use the Levenshtein distance, %
also known as the edit distance, as $d$. %
Because there is no notion of average in the non-Euclidean space $X$ of character sequences with the edit distance, we use the medoid of the $I$ different output transcriptions $y_i$ as our ``mean'' $\hat{y}$.  The medoid $\hat{y}$ of a set $\{y_i\}_{i=1}^I$ is defined as an element of the set whose average distance to all other elements for a distance $d$ is the smallest:
\begin{equation}
\label{eq:medoid}
\hat{y} = \argmin_{y \in \{y_1,\ldots, y_I\}} \sum_{i} d(y, y_i).
\end{equation}
Now that all required notions have been defined, we can define the uncertainty distribution $\mathds{P}^\text{char}_x$ of an audio input $x$ following Eq.~\eqref{eq:uncertainty_distribution}, where $\{y_1, \dots, y_I\}$ are $I$ character sequences output by the ASR engine for different dropout realizations. Note that, as the distances are integers, this distribution is a histogram.

\vspace{-0.1cm}
\subsection{Adversarial ASR Sample Classification}
\label{sec:binary-classification}

We can now classify an input audio sample $x$ as adversarial or not by using a binary classifier taking as input some features derived from the distribution $\mathds{P}^\text{prob}_x$ or $\mathds{P}^\text{char}_x$. In our experiments, we consider the following classifiers: a decision stump trained on the second moment of the distribution (DS), simply comparing that moment to a threshold; a support vector machine (SVM) trained on the first four moments of the distribution (SVM-4); for $\mathds{P}^\text{char}_x$, we also consider an SVM trained on the complete distribution (SVM-F), as we can obtain a fixed-length input vector by considering $(\mathds{P}^\text{char}_x(0),\dots,\mathds{P}^\text{char}_x(C))$, with $C$ set to 19 (no distances larger than 18 were observed on our data); as SVM-F cannot be used for $\mathds{P}^\text{prob}_x$, we replace it with a decision tree trained on the first four moments of the distribution (DecTree).
We use a linear SVM in our experiments; other SVM variants did not provide better results. Note that the DS classifier for $\mathds{P}^\text{prob}_x$ corresponds to the most direct extension of Feinman et al.'s defense to ASR, as mentioned in Section \ref{sec:designing}.

\vspace{-0.1cm}
\section{Experiments and Results}
\label{sec:experiments-and-results}
\vspace{-0.1cm}
We implement our attacks and test our defenses on the Mozilla DeepSpeech \cite{Hannun2014} ASR engine. DeepSpeech is based on a bi-directional RNN network trained with CTC loss.  The model's default dropout rate is $p_\text{tr}\!=\!0.05$ during training.  The adversarial samples are targeted to transcribe as ``okay google unlock phone and delete files''.  In our defense, we use $I\!=\!50$ realizations of dropout to compute the uncertainty of an audio sample. Results are reported in terms of defense accuracy in Table~\ref{table:results} and area under the ROC curve (AUC) in Table~\ref{table:results_AUC}.

\subsection{Choosing a defense dropout rate $p$}
\begin{figure}[t]
	\begin{subfigure}[t]{0.5\columnwidth}
		\centering
        \includegraphics[scale=0.27]{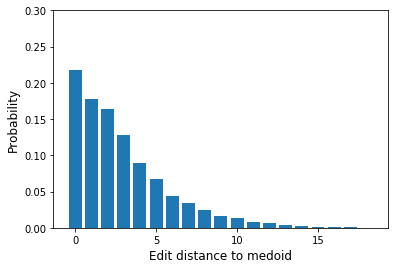}
		\caption{Original}
		\label{fig:original-mean-dis}
	\end{subfigure}%
	\begin{subfigure}[t]{0.5\columnwidth}
		\centering
        \includegraphics[scale=0.27]{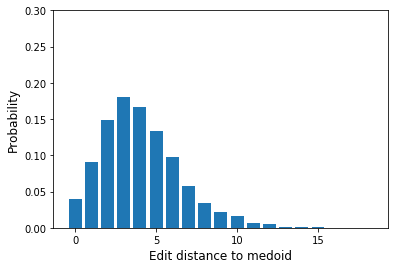}
		\caption{CW, p=0.05}
		\label{fig:CW-005}
	\end{subfigure}%
	\\
	\begin{subfigure}[t]{0.5\columnwidth}
		\centering
        \includegraphics[scale=0.27]{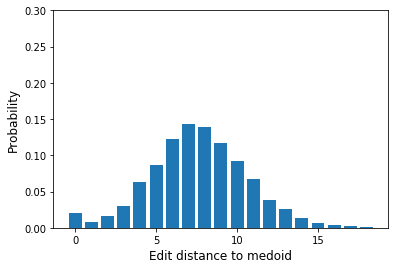}
		\caption{CW, p=0.1}
		\label{fig:CW-01}
	\end{subfigure}%
	\begin{subfigure}[t]{0.5\columnwidth}
		\centering
        \includegraphics[scale=0.27]{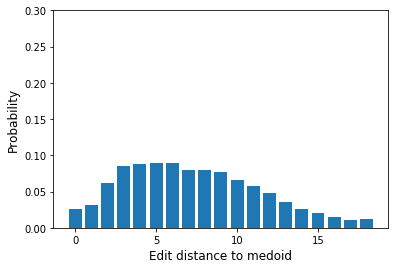}
		\caption{DR, p=0.1}
		\label{fig:DR}
	\end{subfigure}
	\vspace{-0.1cm}
	\caption{Mean %
	uncertainty distribution $\mathds{E}_x[\mathds{P}^\text{char}_x]$ of distances to medoid for original audio samples and adversarial audio samples from different attacks, computed as
	the empirical mean of the distributions over all samples in the training set. The defense dropout rate is denoted by $p$.}
	\label{fig:uncertainty-plots-defense-5}
	\vspace{-0.5cm}
\end{figure}

We first experimented with varying dropout rates
to detect the CW attack samples. While there is little difference between the histograms of original and adversarial samples for $p\!\leq\! 0.04$, we can notice significant differences for $p\!\geq\!0.05$.  Figure~\ref{fig:original-mean-dis} shows the mean uncertainty distribution $\mathds{E}_x[\mathds{P}^\text{char}_x]$ on all original training samples, and Fig.~\ref{fig:CW-005} that on all CW adversarial samples while using a defense dropout rate of 0.05 (CW $p\!=\!0.05$). %

As discussed in Section \ref{sec:dropout-robust}, an adversary may know the defense dropout rate and try training through dropout to break the defense.  We observed that as the dropout rate $p_\text{DR}$ used during adversarial training increases above the default dropout rate $p_\text{tr}=0.05$, there is a sharp decrease in the \textit{forgery success rate} of the attack, i.e., the adversarial sample did not get transcribed as the desired target sentence. An adversary attempting to fool our defense by training through different dropout rates will have a very low forgery success rate for $p_\text{DR}$ above 0.05 and an attack will be nearly impossible for $p_\text{DR}=0.1$.  Hence, we use a dropout rate $p=0.1$ in our defense for all future experiments, and $p_\text{DR}=0.05$ in creating our dropout robust attacks.

\subsection{Results on various attacks}

We trained all classifiers for defense against the DR attack, as it is likely to be the most difficult to detect, and test those models on all attacks (except UrbanSound as it is a separate task). We use a 70-30 train-test split on 500 random samples from the CommonVoice dataset, where each of the original samples is used to generate corresponding adversarial samples for each attack. 
The average duration of each audio sample is about 5 s. %

{\bf CW and DR attacks:} 
The mean uncertainty distributions $\mathds{E}_x[\mathds{P}^\text{char}_x]$
for CW and DR with $p=0.1$ are shown in Figs.~\ref{fig:CW-01} and \ref{fig:DR}. %
We see in Tables~\ref{table:results} and \ref{table:results_AUC} that using a defense dropout rate of 0.1 in conjunction with an SVM trained on the first four moments of the character-sequence-based uncertainty distribution ($\mathds{P}^\text{char}_x$ - SVM-4) %
led to the best results on these attacks. %

{\bf Noise reduction robust (NRR) attack:} 
We observed that the NRR attacks were indeed robust to denoising techniques, and despite having been optimized through a spectral subtraction algorithm, our adversarial examples were also fairly robust to other noise reduction algorithms such as the \texttt{logmmse} algorithm \cite{logmmse}.
From Table \ref{table:results}, we see that the accuracy results for $\mathds{P}^\text{char}_x$ on NRR and DR are similar, which may be explained by the fact that they are both designed to be robust to perturbations. %

{\bf Imperceptible audio attack (IA):} 
The IA attack was originally implemented to work with attention-based models.  We re-implemented it for the Mozilla DeepSpeech engine with a few modifications: the learning rate for the initial $\alpha\!=\!0$ stage was decreased from 100 to 10, and the learning rate for the $\alpha\!>\!0$ stages was decreased from 1 to 0.1; %
furthermore, the loss function used is CTC instead of cross-entropy loss. The IA attacks are audibly sharper and cleaner compared to the vanilla CW attack. Their behavior against our defense is however similar to CW in terms of accuracy. %
We also implemented the IA attack on samples from the LibriSpeech \cite{librispeech} dataset to investigate performance on longer utterances than the CommonVoice dataset. The extra length resulting in longer computation times to create adversarial examples, we only evaluated 20 samples. Our defense was able to detect all samples without error.

\begin{table}[!t]
	\caption{Detection accuracy [\%] on various attacks for the different classifiers. $p$ denotes the defense dropout rate. %
	}\vspace{-0.2cm}
	\label{table:results} 
	\centering
		\small
		\resizebox{\columnwidth}{!}{\setlength{\tabcolsep}{3pt}%
		\begin{tabular}{c|c c c c c c c}
			\toprule
			\multicolumn{2}{c}{} & $p=0.05$ & \multicolumn{5}{c}{$p=0.1$} \\
			\cmidrule(lr){3-3} 			\cmidrule(lr){4-8}
			\multicolumn{2}{c}{} & CW  & CW & DR & NRR & IA & US\\
			\midrule
			&DS &  71.7 & 83.3 & 82.5 & 75.5 & 91.0 & 90.4\\
		$\mathds{P}^\text{prob}_x$ &SVM-4 & 66.7 & 80.8 & 68.0 & 53.3 & 68.0 & 64.4\\
			&DecTree  & 65.0 & 80.8 & 72.0 & 70.0 & 73.3 & 91.8\\			\midrule %
			&DS & 72.3 & \bfseries 96.5 & 81.0 & 81.0 & \bfseries 92.0 & 79.0\\
	$\mathds{P}^\text{char}_x$	&SVM-4 & 76.7 & \bfseries 96.5 & \bfseries 88.5 & \bfseries 88.5 & \bfseries 92.0 &               \bfseries 93.9\\
			&SVM-F  & 74.0 & 85.8 & 86.5 & 87.5 & 88.3 & 83.0\\
			\bottomrule
		\end{tabular}
		}\vspace{-0.1cm}
\end{table}

\begin{table}[t]
	\caption{AUC score on various attacks for the different classifiers. $p$ denotes the defense dropout rate. %
	}\vspace{-0.2cm}
	\label{table:results_AUC} 
	\centering
		\small
		\resizebox{\columnwidth}{!}{\setlength{\tabcolsep}{3pt}%
		\begin{tabular}{c|c c c c c c c}
			\toprule
			\multicolumn{2}{c}{} & $p=0.05$ & \multicolumn{5}{c}{$p=0.1$} \\
			\cmidrule(lr){3-3} 			\cmidrule(lr){4-8}
			\multicolumn{2}{c}{} & CW  & CW & DR & NRR & IA & US\\
			\midrule
			&DS & 0.72 & 0.85 & 0.83 & 0.84 & 0.82 & 0.91\\
		$\mathds{P}^\text{prob}_x$ &SVM-4 & 0.84 & 0.91 &  0.88 & 0.89 & 0.90 & \bfseries 0.98\\
			&DecTree  & 0.72 & 0.85 & 0.83 & 0.84 & 0.82 & 0.91\\			\midrule %
			&DS & 0.72 & 0.82 & 0.81 & 0.82 & 0.73 & 0.86\\
	$\mathds{P}^\text{char}_x$	&SVM-4 & \bfseries 0.88 & \bfseries 0.92 & \bfseries 0.95 & \bfseries 0.93 & \bfseries 0.95 &  0.94\\
			&SVM-F  & 0.75 & 0.91 & 0.92 & 0.93 & 0.94 & 0.74\\
			\bottomrule
		\end{tabular}
		}%
\vspace{-1em}
\end{table}

{\bf Urban sound attack:} 
We were able to successfully apply the vanilla CW attack to the UrbanSound (US) \cite{Salamon:UrbanSound:ACMMM:14} dataset.  Unlike the previous results, the mean distribution of the original samples did not resemble Fig.~\ref{fig:original-mean-dis} as the input was no longer speech, but the mean distribution of the adversarial samples did resemble Fig.~\ref{fig:DR} despite not being trained through dropout. The classifier was trained on a similar dataset as that used to train the DR defense, but based on data from UrbanSound instead of CommonVoice. The results are shown in Tables \ref{table:results} and  \ref{table:results_AUC}.

{\bf Universal Perturbation:} 
The universally perturbed audio attacks proposed in \cite{Neekhara2019} do not fall under our definition of an attack as the adversarial example often does not transcribe as a meaningful sentence and hence has no malicious nature. Nevertheless, our $\mathds{P}^\text{char}_x$ - SVM-4 defense trained on the CW attack data was able to detect the adversarial examples on the authors' website with 100\% accuracy.

{\bf Entropy as a measure of uncertainty:}  After this article was submitted, concurrent work exploring various options for detecting audio attacks was released on arXiv \cite{daubener2020detecting}. That work included a method based on dropout and the Feinman-like variance similar to our $\mathds{P}^\text{prob}_x$ - DS method, and reported obtaining better results with entropy. In preliminary experiments, we found that our $\mathds{P}^\text{char}_x$ - SVM-4 defense performed similarly to or better than an entropy-based method in terms of accuracy (e.g., 96.5 \% vs 90.5 \% on CW, 88.5 \% vs 88.0 \% on DR), and significantly better in terms of AUC (e.g., 0.92 vs 0.81 on CW, 0.95 vs 0.88 on DR). The entropy feature may also potentially be combined with the features derived using our method. A more thorough comparison will be the object of future work.

\section{Conclusion}
\label{sec:conclusion}

In this paper, we showed that it is possible to extend the vanilla CW attack to create adversarial examples robust to dropout and denoising, and that such attacks can also be embedded within everyday urban sounds. We developed a defense that can detect a wide range of attacks on ASR engines by leveraging the uncertainty introduced by dropout. Using simple classifiers, we can detect adversarial examples with high confidence.  Particularly, training an SVM on the first four moments of the distributions of distances between character sequences realized with dropout and their medoid achieves the best results. Finally, our defense is able to detect adversarial examples obtained with frequency masking or with a model based on universal perturbations.

\vfill\pagebreak
\balance

\bibliographystyle{IEEEtran}

\bibliography{refs}

\end{document}